\documentclass[preprint]{revtex4-1}
\usepackage[english]{babel}
\usepackage[utf8]{inputenc}
\usepackage{hyperref}
\usepackage{amssymb}
\usepackage{amsthm} 
\usepackage{amsmath}
\usepackage{graphicx}
\usepackage{tikz}
\usepackage{color}
\usepackage{epsfig}
\usepackage{verbatim}
\usepackage{pgf}
\usepackage{pstricks}
\usepackage{natbib}

\newcommand{\mc}[1]{\mathcal{#1}}
\newcommand{\avg}[1]{\langle#1\rangle}
\newcommand{\Avg}[1]{[#1]}
\newcommand{\aAvg}[3]{\langle#1|#2|#3]}

\newcommand{\MHV}{\mathrm{MHV}}
\newcommand{\NMHV}{\mathrm{NMHV}}
\newcommand{\NkMHV}{\mathrm{N}^k\mathrm{MHV}}
\newcommand{\gMHV}{\overline{\mathrm{MHV}}}

\begin{document}
\date{\today}
\author{Mads Søgaard}
\affiliation{Niels Bohr International Academy and Discovery Center, 
Niels Bohr Institute, Blegdamsvej 17, DK-2100 Copenhagen, Denmark}
\title{Supersums for all supersymmetric amplitudes}

\begin{abstract}
We present an on-shell graphical framework for superamplitudes in super 
Yang-Mills theory with arbitrary supersymmetry. Our diagrammatic procedure is
derived through manipulations of the full $\mc N = 4$ superamplitude and 
illustrated by a number of explicit examples.
\end{abstract}

\maketitle

\section{Introduction}
Multiloop scattering amplitudes in maximally supersymmetric ($\mc N = 4$)
Yang-Mills theory have been studied extensively over the years
\cite{Bern:2009xq,Elvang:2011fx,Bern:2006ew,Anastasiou:2003kj,Bern:2005iz,
Cachazo:2006tj,Drummond:2007au,Bern:2007ct,Drummond:2009fd,Berkovits:2008ic,
Bartels:2008ce,Bern:2008ap} in connection with for instance the famous AdS/CFT 
correspondence and possible finiteness of supergravity theories. Remarkable 
results have been uncovered including new favorable evaluation methods 
applicable to both tree- and loop-level amplitudes.

An essential part of this progress is the on-shell superspace formalism, which 
organizes on-shell states and scattering amplitudes in maximally supersymmetric 
super Yang-Mills theory very elegantly \cite{Bern:2009xq,Nair:1988bq,
Bianchi:2008pu,Elvang:2008na,Georgiou:2004by,Elvang:2008vz,Elvang:2011fx,
ArkaniHamed:2008gz,Drummond:2008vq,Brandhuber:2008pf,Drummond:2008cr,
Drummond:2008bq}. The principle is to arrange the entire supermultiplet as a 
convenient expansion labeled by $R$-symmetry indices and particle number into
$n$ superfields, one for each external leg. All possible scattering combinations
are realized by formation of superamplitudes, defined as generating functions
with the superfields as input, having all supersymmetric Ward identities
automatically satisfied. Individual scattering amplitudes are available from the
generating function using appropriate combinations of Grassmann differential
operators.  Using either the Maximally Helicity Violating (MHV) vertex expansion
\cite{Cachazo:2004kj,Risager:2005vk} or the Britto-Cachazo-Feng-Witten (BCFW)
on-shell recursion relations \cite{Britto:2004ap,Britto:2005fq} superamplitudes
for general particle and helicity configurations may be constructed.

In a recent paper \cite{Elvang:2011fx}, overlapping with \cite{Lal:2009gn}, the
maximally supersymmetric superspace setup was generalized to super Yang-Mills
theory with reduced supersymmetry, i.e. with $\mc N < 4$ generators of
supersymmetry. All necessary steps towards developing both a holomorphic and a
non-holomorphic approach were taken. In particular, it was shown that the most
general MHV generating function valid for generic supersymmetry can be derived
by simple combinations of truncations and Fourier transforms of the full 
$\mc N = 4$ superamplitude.

The introduction of $\mc N = 4$ superamplitudes spawned important developments
including a very convenient diagrammatic representation \cite{Bern:2009xq}. For
brevity, all possible contractions between external states are tracked, yielding 
a one-to-one correspondence between diagrams and individual scattering 
amplitudes. Motivated by \cite{Elvang:2011fx} we will extend this graphical 
framework to $\mc N < 4$ superamplitudes.
\newpage

\section{$\mc N = 4$ Superamplitudes}
Before developing the superspace formalism and discussing the MHV and $\gMHV$ 
generating functions for super Yang-Mills theory with less than maximal 
supersymmetry, we briefly remind ourselves about the $\mc N = 4$ setup.

The $\mc N = 4$ vector multiplet is uniquely CPT self-conjugate allowing all 
on-shell states to be incorporated into a single holomorphic superfield 
$\Phi(p,\eta)$ written as an expansion in Grassmann variables $\eta_a$ 
with $a = 1,\dots,4$ being $R$-symmetry indices. Within this framework the sixteen 
physical states in the $\mc N = 4$ supermultiplet are two gluons $g_+$ and 
$g^{abcd}_-$, four gluino pairs $f^a_+$ and $f^{abc}_-$, plus six real scalars 
$s^{ab}$, all completely antisymmetric. The superfield then takes the form 
\cite{Drummond:2008cr} 
\begin{align}
\label{N=4SUPERFIELD}
\Phi(p,\eta) = g_++\eta_af^a_++\frac{1}{2!}\eta_a\eta_bs^{ab}
               +\frac{1}{3!}\eta_a\eta_b\eta_c f^{abc}_-
               +\frac{1}{4!}\eta_a\eta_b\eta_c\eta_dg^{abcd}_-\;.
\end{align}
Grassmann Fourier transformation yields an antiholomorphic superfield,
\begin{align}
\label{N=4ANTISUPERFIELD}
\tilde\Phi(p,\tilde\eta) = 
g^-+\tilde\eta^af_a^-+\frac{1}{2!}\tilde\eta^a\tilde\eta^bs_{ab}
+\frac{1}{3!}\tilde\eta^a\tilde\eta^b \tilde\eta^c f^+_{abc}
+\frac{1}{4!}\tilde\eta^a\tilde\eta^b \tilde\eta^c\tilde\eta^dg^+_{abcd}\;,
\end{align}
but with the exact same particle content encoded. Because of this equivalence 
either representation may be preferred. 

In order to shed light on how proliferation of amplitudes in $\mc N = 4$ super
Yang-Mills theory is handled, the concept of superamplitudes is introduced. 
With $n$ copies of the superfield $\Phi_i$ arranged, we organize the full 
$n$-point tree-level superamplitude ascendingly according to Grassmann degree 
in steps of four,
\begin{align}
\mc A_n(p,\eta) = \mc A(\Phi_1\cdots\Phi_n) = 
\mc A_n^{\MHV}+\mc A_n^{\NMHV}+\cdots+\mc A_n^{\gMHV}\;,
\end{align}
ranging from eight $\eta$'s to $4n-8$. Explicit formulas for all $\NkMHV$
amplitudes relying on BCFW shifts
\cite{Britto:2004ap,Britto:2005fq,Brandhuber:2008pf,ArkaniHamed:2008gz}
exist in the litterature \cite{Drummond:2008cr}, but our focus is on $\MHV$ and 
$\gMHV$ amplitudes. It follows that all MHV amplitudes may be collected into a 
generating function, which we will call the MHV superamplitude, such that each 
term corresponds to a regular scattering amplitude involving gluons, fermions and 
scalars. The MHV superamplitude is defined as \cite{Bern:2009xq}
\begin{align}
\mc A^{\mathrm{MHV}}_{n}(1,2,\dots,n) = 
i\frac{(2\pi)^4\delta^{(4)}(\sum_{j=1}^n p_j)}{\prod_{m=1}^{n}\avg{m(m+1)}}
\delta^{(8)}\left(\sum_{j=1}^n\lambda_j^\alpha\eta_{ja}\right)\;,
\end{align}
and contains in addition to the well-known overall momentum conservation,
an eightfold Grassmann delta function, which conserves supermomentum
$Q^\alpha_a\equiv \sum_{j=1}^n\lambda_j^\alpha\eta_{ja}$. It proves
advantageous to expand the superamplitude as a sum of monomials in the 
$\eta$'s. Factorization in the group index
and $\delta(\eta) = \eta$ for Grassmann variables imply that
\begin{align}
\delta^{(8)}\left(\sum_{j=1}^n\lambda_j^\alpha\eta_{ja}\right) =
\prod_{a=1}^4\delta^{(2)}\left(\sum_{j=1}^n\lambda_j^\alpha\eta_{ja}\right) =
\prod_{a=1}^4\sum_{i<j}\avg{ij}\eta_{ia}\eta_{ja}\;.
\end{align}
Consequently, the MHV superamplitude can be recast as
\begin{align}
\label{MHVSUPERAMPLITUDE}
\mc A^{\mathrm{MHV}}_n(1,2,\dots,n) = 
i\frac{\prod_{a=1}^4\sum_{i<j}\avg{ij}\eta_{ia}\eta_{ja}}
{\prod_{m=1}^{n}\avg{m(m+1)}}\;,
\end{align}
with four-momentum conservation stripped.

Notation of component amplitudes is streamlined in terms of spinor products of 
supermomenta of the individual legs defined by
\begin{align}
\avg{q_{ia}q_{ja}}\equiv \eta_{ia}\avg{ij}\eta_{ja}\;, \qquad
\Avg{\tilde q_i^a\tilde q_j^a}\equiv \tilde\eta_i^a\Avg{ij}\tilde\eta_j^a\;.
\end{align}
For four external legs, some simple examples of component amplitudes are
\begin{align}
\label{SIMPLEMHV1}
A_4^{\mathrm{tree}}(
1^-_{g^{1234}},2^-_{g^{1234}},3^+_{g},4^+_{g}) &=
i\frac{\prod_{a=1}^4\avg{q_{1a}q_{2a}}}{\avg{12}\avg{23}\avg{34}\avg{41}}\;, \\
\label{SIMPLEMHV2}
A_4^{\mathrm{tree}}(
1^-_{g^{abcd}},2^-_{f^{abc}},3^+_{f^d},4^+_{g}) &=
i\frac{\avg{q_{1a}q_{2a}}\avg{q_{1b}q_{2b}}\avg{q_{1c}q_{2c}}\avg{q_{1d}q_{3d}}}
{\avg{12}\avg{23}\avg{34}\avg{41}}\;, \\
A_4^{\mathrm{tree}}(
\label{SIMPLEMHV3}
1^-_{f^{abc}},2^-_{f^{abd}},3_{s^{cd}},4^+_{g}) &=
i\frac{\avg{q_{1a}q_{2a}}\avg{q_{1b}q_{2b}}\avg{q_{1c}q_{3c}}\avg{q_{2d}q_{3d}}}
{\avg{12}\avg{23}\avg{34}\avg{41}}\;.
\end{align}

Analogous to the MHV superamplitude one can define the $\gMHV$ superamplitude
\cite{Bern:2009xq} by
\begin{align}
\mc A^{\mathrm{\gMHV}}_n(1,2,\dots,n) &= 
i(-1)^n\frac{\delta^{(8)}(\sum_{j=1}^n\tilde\lambda_{j\dot\alpha}\tilde\eta_j^a)}
{\prod_{m=1}^{n}\Avg{m(m+1)}} =
i(-1)^n\frac{\prod_{a=1}^4\sum_{i<j}^n\Avg{\tilde q_i^a\tilde q_j^a}}
{\prod_{m=1}^{n}\Avg{m(m+1)}}\;,
\end{align}
built entirely from antiholomorphic superfields \eqref{N=4ANTISUPERFIELD}. It is
mapped from the $\tilde\eta$ coordinates to the untilded superspace using the
Grassmann Fourier transform.

\section{All MHV Superamplitudes}
It is desirable to extend the neat $\mc N = 4$ superspace formulation to
super Yang-Mills theory with nonmaximal supersymmetry. Two different approaches
exist \cite{Elvang:2011fx}, neither of which can be described in terms of a 
single superfield, due to the fact that $\mc N < 4$ super Yang-Mills theory is 
not CPT self-conjugate. Instead the physical states of the $\mc N = 1,2,3$
supermultiplets can be assembled into either two conjugate superfields
$\Phi^{\mc N}$ and $\tilde\Phi^{\mc N}$, which are related to 
\eqref{N=4SUPERFIELD} and \eqref{N=4ANTISUPERFIELD} by the truncations 
$\eta_{\mc N +1,\dots,4}\to 0$ and $\tilde\eta^{\mc N+1,\dots,4}\to 0$ 
respectively, or two holomorphic superfields $\Phi^{\mc N}$ and $\Psi^{\mc N}$ 
obtained from the $\mc N = 4$ superfield by suitable combinations of 
truncations and Grassmann integrations.

Here we focus on the $\Phi-\Psi$ formalism, since we would like to avoid the
unfortunate mixing of the $\eta$ and $\tilde\eta$ variables in the 
$\Phi-\Phi^\dagger$ picture, which does not lead to any obvious graphical 
interpretation.

Consider, say, the $\mc N = 1$ conjugate superfields obtained by letting 
$\eta_{2,3,4}\to 0$ and $\tilde\eta^{2,3,4}\to 0$ in \eqref{N=4SUPERFIELD}
and \eqref{N=4ANTISUPERFIELD}. It follows that $\Phi^{\mc N = 1}(p,\eta) =
g_++\eta_af^a_+$ and $\tilde\Phi^{\mc N = 1}(p,\tilde\eta) =
g^-+\tilde\eta^af_a^-$. The $n$-point MHV configuration then has two states from
the $\tilde\Phi^{\mc N = 1}$ superfield and $n-2$ from the other. The essense of 
the $\Phi-\Psi$ approach is to keep the truncated holomorphic superfields, but 
discard the ones of negative overall helicity. To achieve this we observe that 
full description of the particle content can be maintained by supplementing 
$\Phi_i$ by a new holomorphic superfield introduced in \cite{Elvang:2011fx},
\begin{align}
\label{PUREMHVTRUNCINT}
\Psi_i^{\mc N}(p,\eta) &\equiv 
\int\left(\prod_{a=\mc N+1}^4d\eta_{ia}\right)\Phi_i^{\mc N = 4}(p,\eta)\;,
\end{align}
where for instance $\Psi^{\mc N = 1} = f_-+\eta_a g_-^a$ with $a = 1$ fixed but
kept for notational uniformity.

For clarity we list here all pairs of superfields $\left[\Phi^{\mc N},\Psi^{\mc
N}\right]$ for $\mc N = 1,2,3$. States of the $\Psi$-sector are hatted to be
distinguished from those originating from the non-manipulated superfields. In
this notation the indices take values $a = 1,\dots,\mc N$.
\begin{align}
\Phi^{\mc N = 1} & = g_++\eta_af^a_+ \; , \nonumber \\
\Psi^{\mc N = 1} & = \hat{f}_-+\eta_a\hat{g}_-^a \; , \\
\Phi^{\mc N = 2} & =
g_++\eta_af^a_++\frac{1}{2!}\eta_a\eta_bs^{ab} \; , \nonumber \\
\Psi^{\mc N = 2} & =
\hat{s}+\eta_a\hat{f}^a_-+\frac{1}{2!}\eta_a\eta_b\hat{g}^{ab}_- \; , \\
\Phi^{\mc N = 3} & =
g_++\eta_af^a_++\frac{1}{2!}\eta_a\eta_bs^{ab}
+\frac{1}{3!}\eta_a\eta_b\eta_c f^{abc}_- \; , \nonumber \\
\Psi^{\mc N = 3} & = \hat{f}_++\eta_a\hat{s}^a
+\frac{1}{2!}\eta_a\eta_b\hat{f}^{ab}_-
+\frac{1}{3!}\eta_a\eta_b\eta_c\hat{g}^{abc}_- \; .
\end{align}

Let us define what is understood by a MHV amplitude in $\mc N < 4$ super
Yang-Mills theory in this notation. First of all, a MHV amplitude must have 
$2\mc N$ paired indices as a consequence of truncation of the superfields. 
Moreover, to be MHV requires two states from the $\Psi$ sector and $n-2$ from 
the $\Phi$ superfield. With this in mind we are now ready to derive the MHV 
superamplitude for $\mc N < 4$. Recall that the maximally supersymmetric 
superamplitude is a function of $\Phi_i$, $i = 1,\dots,n$. Suppose legs $i$ and 
$j$ represent states of the $\Psi$-sector. We can then convert $\Phi_i$ and 
$\Phi_j$ to $\Psi_i$ and $\Psi_j$ by integrating out $(4-\mc N)$ $\eta_i$'s 
and $\eta_j$'s according to \eqref{PUREMHVTRUNCINT}. Afterwards we truncate 
the remaining $n-2$ legs to reduce the content of the $\Phi$ superfields. 
The only obstacle is to rearrange the integration measure appropriately, but 
actually the overall sign is not very important to us. 
\begin{align}
\label{MHVNLESSSUPERAMPLITUDE}
\mc A^{\mc N,\; \MHV}_{n,ij} & = 
i\int\left(\prod_{b\,=\,\mc N+1}^4d\eta_{ib}
\prod_{c\,=\,\mc N+1}^4d\eta_{jc}\right)
\frac{\prod_{a=1}^4\sum_{k<l}^n\avg{kl}\eta_{ka}\eta_{la}}
{\prod_{m=1}^{n}\avg{m(m+1)}}
\bigg|_{\mathrm{truncate}} \nonumber \\ & =
\frac{i(-1)^{\frac{1}{2}\mc N(\mc N-1)}}
{\prod_{m=1}^{n}\avg{m(m+1)}}
\int\left(\prod_{b\,=\,\mc N+1}^4d\eta_{jb} d\eta_{ib}\right)
\prod_{a=1}^4\sum_{k<l}^n\avg{kl}\eta_{ka}\eta_{la}
\bigg|_{\mathrm{truncate}} \nonumber \\ & =
i(-1)^{\frac{1}{2}\mc N(\mc N-1)}
\frac{\avg{ij}^{4-\mc N}\prod_{a=1}^{\mc N}\sum_{k<l}
\avg{kl}\eta_{ka}\eta_{la}}
{\prod_{m=1}^{n}\avg{m(m+1)}}\;.
\end{align}
It is noticed that this generic MHV superamplitude, which is equivalent to the
one derived in \cite{Bern:2009xq}, as expected reduces to the original 
$\mc N = 4$ MHV superamplitude for maximal supersymmetry, and to the
Parke-Taylor formula for pure Yang-Mills theory ($\mc N = 0$).

\section{Diagrammatic Representation}
It would be useful to have a simple visualization of the rather abstract
expression for the $\mc N \leq 4$ superamplitude. Similar to Feynman graphs,
such a scheme should yield a one-to-one correspondence between diagrams and 
component amplitudes.

The basic ingredients of the $\mc N = 4$ MHV superamplitude appear in
\eqref{MHVSUPERAMPLITUDE} as a cyclic spinor string in the denominator and
spinor products of supermomenta in the numerator. The pictorial representation
for these components was developed in \cite{Bern:2009xq} and the transition to 
$\mc N < 4$ was sketched. Taking \eqref{MHVNLESSSUPERAMPLITUDE} into account we
see that the only structural difference between the maximally supersymmetric and
the $\mc N < 4$ generating function is the number of spinor products of
supermomenta. Contrary to our $\mc N < 4$ MHV superamplitude, the 
$4-\mc N$ integrated index lines responsible for the $\avg{ij}^{4-\mc N}$ factor
still carry Grassmann variables with fixed $R$-symmetry indices in
\cite{Bern:2009xq}. We circumvent these by introducing a new {\it sector line} 
between the two states of the $\Psi$-sector to catch the $\avg{ij}^{4-\mc N}$ 
factor of \eqref{MHVNLESSSUPERAMPLITUDE}.

\begin{figure}[!h]
\centering
\includegraphics[scale=0.7]{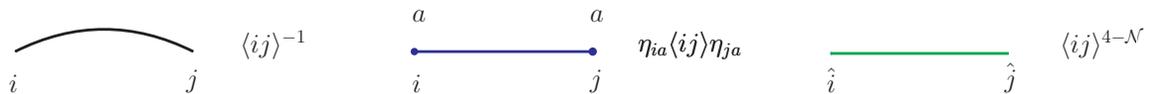}
\caption{\label{INDEXRULES}
The individual terms in the spinor string in the denominator are 
diagrammatrically represented by a curved line without endpoints. The blue index
line with endpoints translates into a spinor product of supermomenta of the
corresponding individual legs. For $\mc N < 4$, states of the $\Psi$-sector must
be connected with a solid green sector line without end points. Identical graphs 
exist in the $\gMHV$ picture with obvious continued expressions. }
\end{figure}

In order to construct an index diagram for a given MHV amplitude with $n$
external legs in $\mc N \leq 4$ super Yang-Mills theory having $2\mc N$ paired 
$R$-symmetry indices, simply follow this prescription:
\begin{enumerate}
\setlength{\itemsep}{1pt}
\setlength{\parskip}{0pt}
\setlength{\parsep}{0pt}
\item Draw a polygon with $n$ sides of solid, black lines curving inwards,
leaving space at each corner for external legs.
\item Distribute the external legs at the gaps and label them with 
appropriate momentum, helicity and $R$-symmetry indices.
\item Connect paired $R$-symmetry indices with blue index lines with
endpoints.
\item For $\mc N < 4$ insert a single solid green sector line without endpoints 
between the two states of the $\Psi$-sector.
\item Indicate holomorphicity with $\oplus$ for MHV and $\ominus$ for 
$\gMHV$ in the center of the diagram.
\end{enumerate}
The diagram rules are summarized in Fig.~\ref{INDEXRULES}.

\section{Tree-level examples}
\begin{figure}[!h]
\centering
\includegraphics[scale=0.6125]{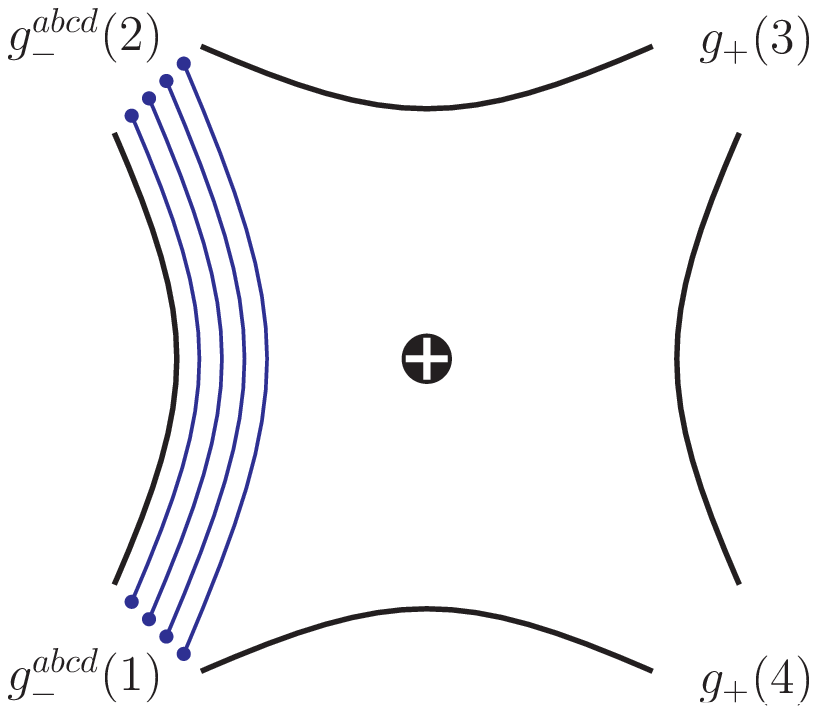}
\includegraphics[scale=0.6125]{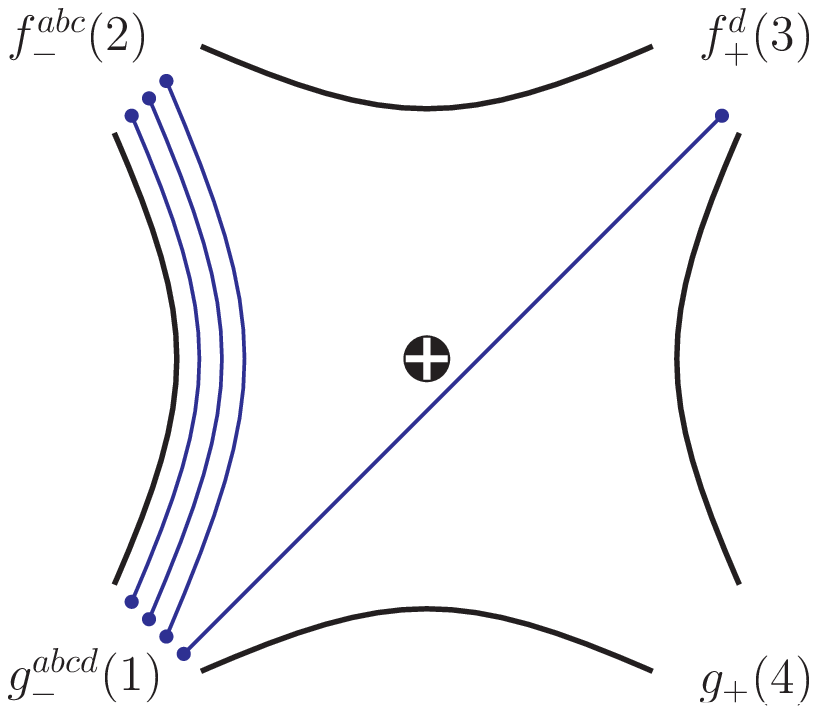}
\includegraphics[scale=0.6125]{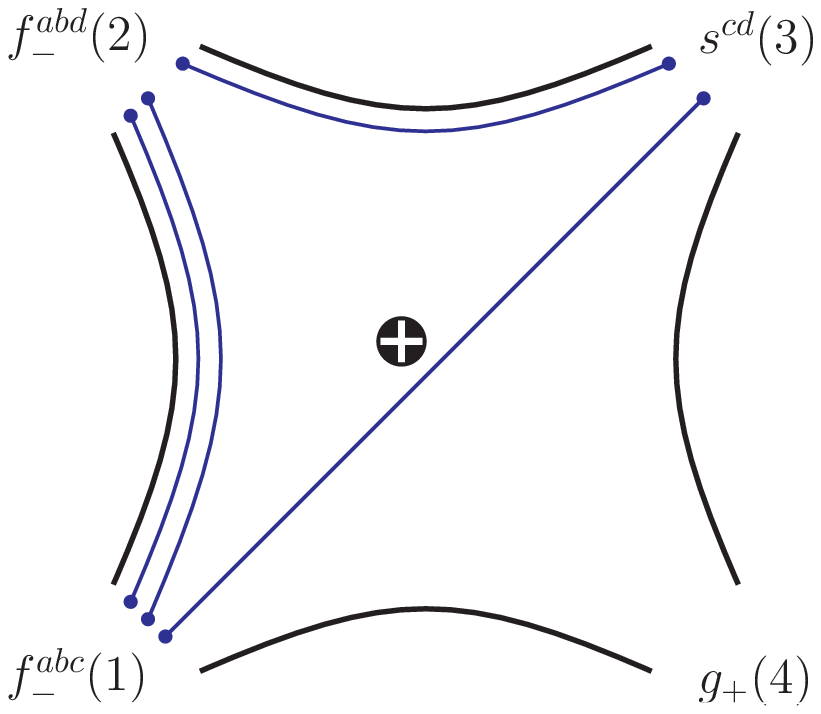}
\caption{\label{N=4INDEXDIAGRAMS}
The analytic expressions for the $\mc N = 4$ amplitudes
\eqref{SIMPLEMHV1}-\eqref{SIMPLEMHV3} are neatly captured by these three simple
index diagrams. If reinterpreted in the $\gMHV$ picture the first diagram would 
have four index lines between the two positive helicity gluons for instance.}
\end{figure}

\begin{figure}[!h]
\centering
\includegraphics[scale=0.6125]{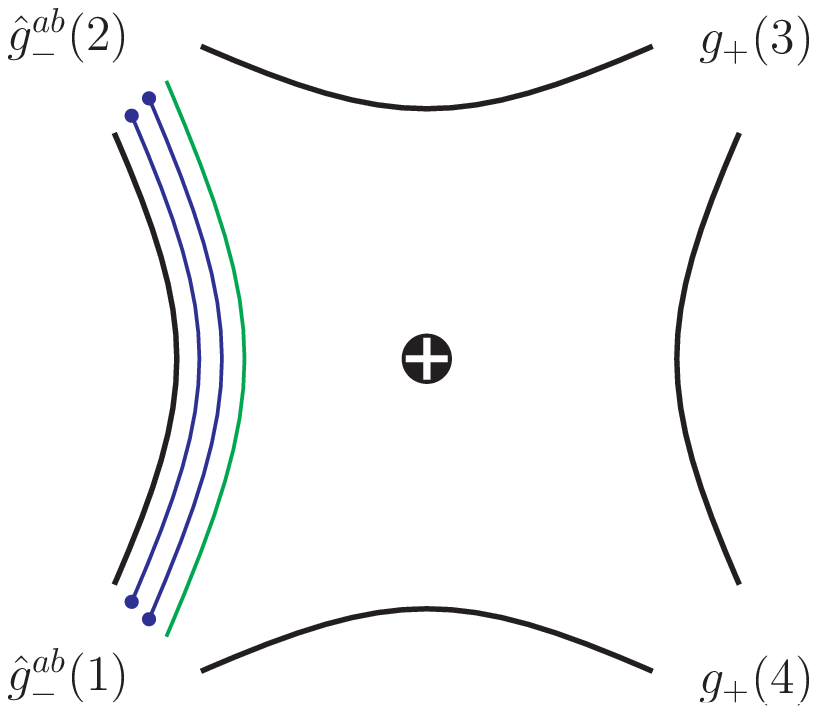}
\includegraphics[scale=0.6125]{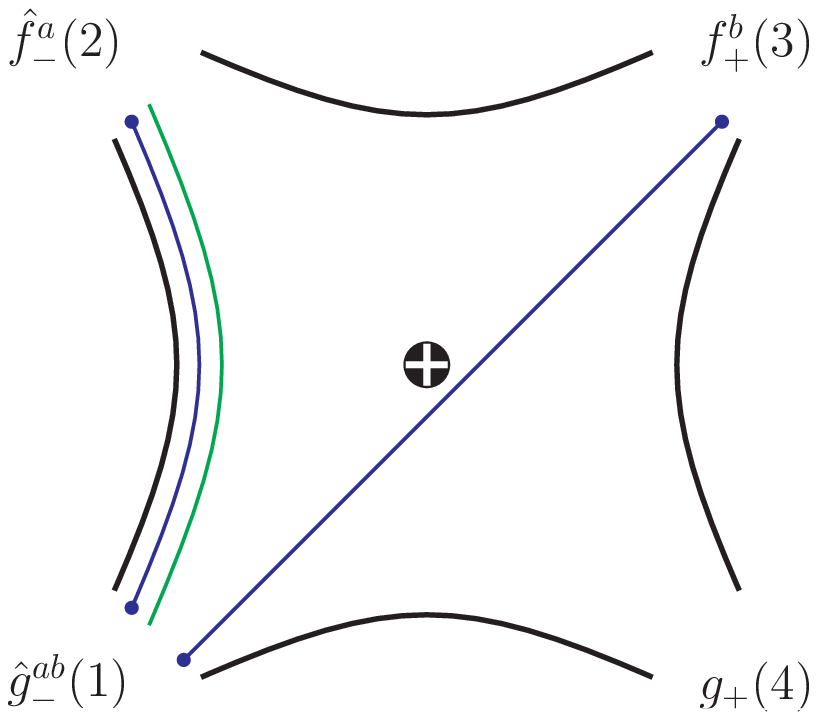}
\includegraphics[scale=0.6125]{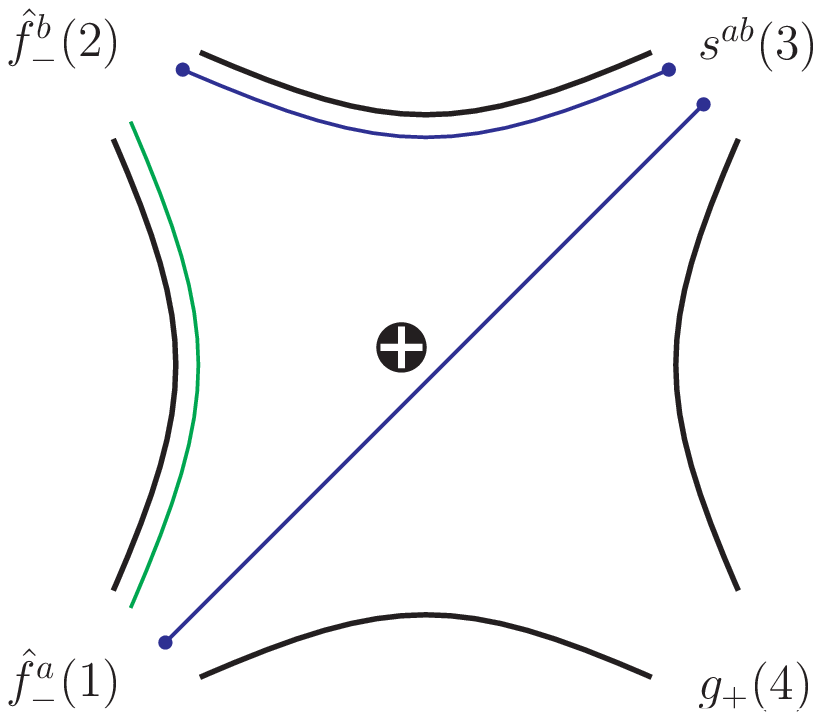}
\caption{\label{N=2INDEXDIAGRAMS}
Many amplitudes exist for several values of $\mc N$. Here 
\eqref{SIMPLEMHV1}-\eqref{SIMPLEMHV3} are shown for $\mc N = 2$. In general, if
a diagram has at most $\Lambda$ grouped index lines in $\mc N = 4$, then it can
be non-zero for reduced supersymmetry only provided $\Lambda \geq 4-\mc N$.}
\end{figure}

Let us now see how the diagrammatic representation scheme works out in practice 
for a number of MHV amplitudes at tree-level in super Yang-Mills theory with
$\mc N = 4$ and fewer supersymmetries. We start with the amplitudes expressions 
\eqref{SIMPLEMHV1}-\eqref{SIMPLEMHV3}, whose corresponding diagrams are drawn 
Fig.~\ref{N=4INDEXDIAGRAMS}. These amplitudes are viewed for maximal 
supersymmetry without sector lines, but actually they exist for all $\mc N$, 
$\mc N\geq 1$ and $\mc N\geq 2$ respectively. In Fig.~\ref{N=2INDEXDIAGRAMS} 
the corresponding $\mc N = 2$ graphs are given. Thanks to the $\avg{12}^2$ 
factor from the sector line the diagrams of Figs.~\ref{N=4INDEXDIAGRAMS} 
and~\ref{N=2INDEXDIAGRAMS} have pairwise identical expressions.
% MHV N = 4 graphs

% MHV N = 2 graphs
\begin{figure}[!h]
\centering
\includegraphics[scale=0.6125]{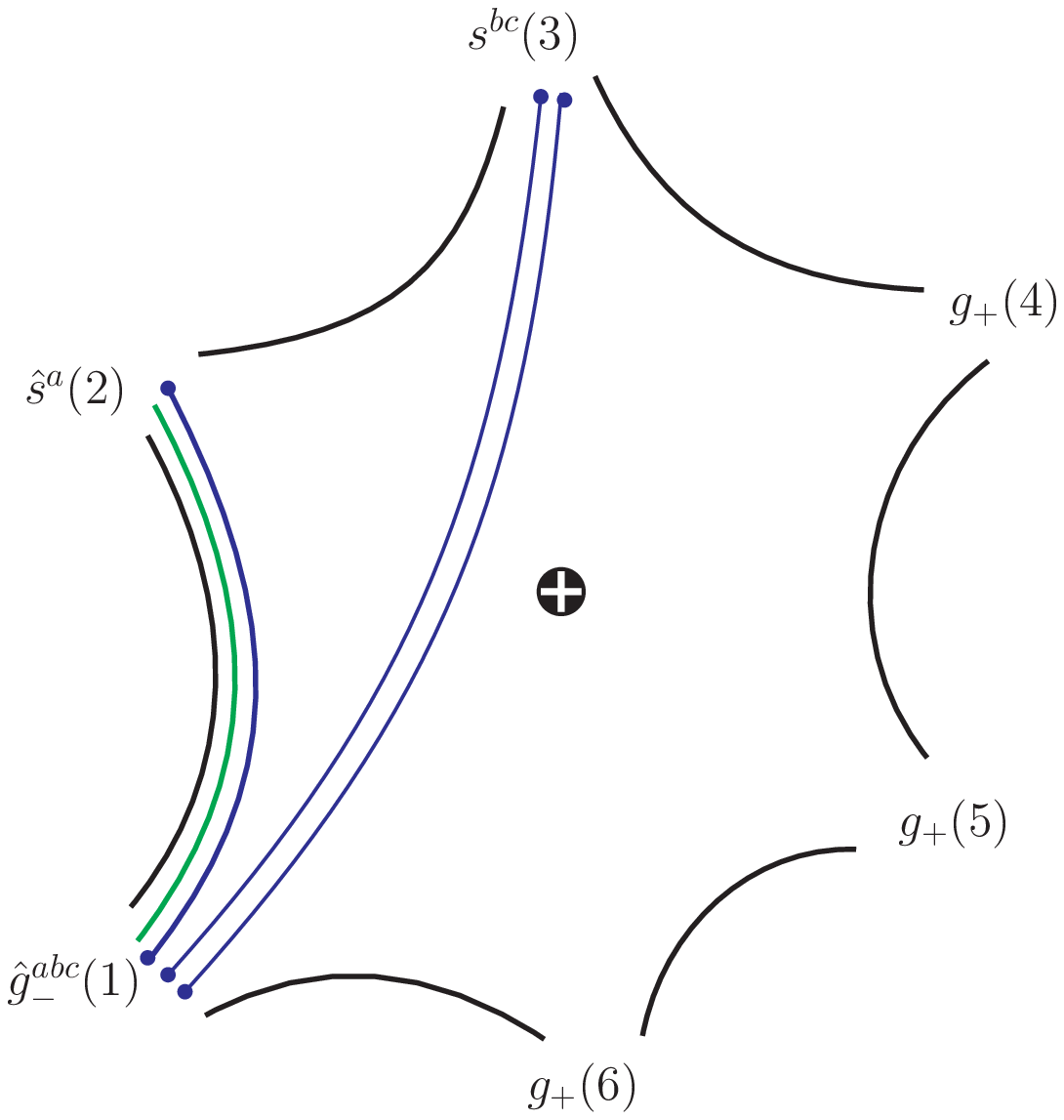}
\includegraphics[scale=0.6125]{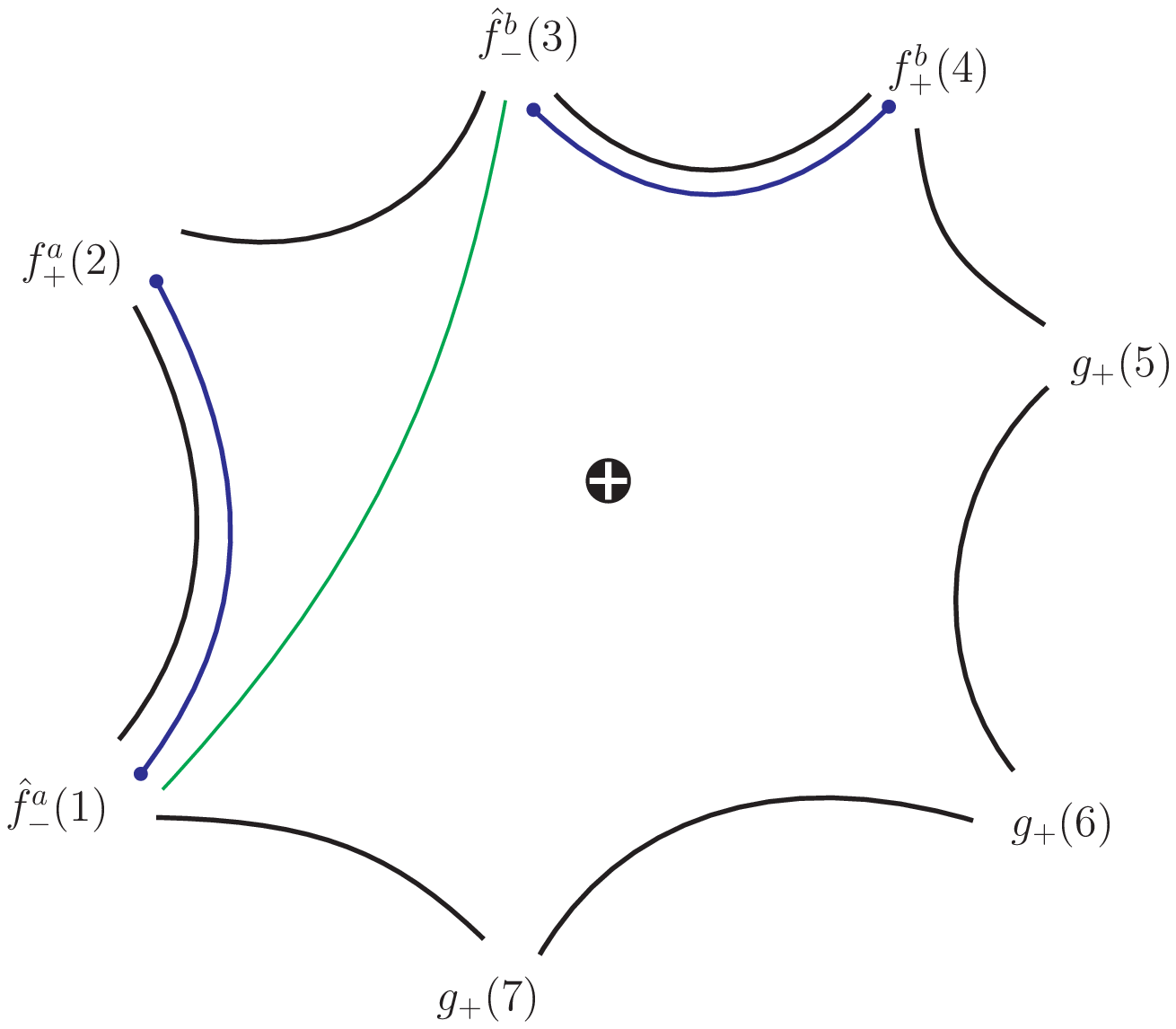}
\caption{\label{SIXSEVENPOINT}
$\mc N = 3$ and $\mc N = 2$ MHV tree-level examples as suggested by the 
number of index lines.
}
\end{figure}

\newpage
We have now shown how easy it is to draw tree-level diagrams for MHV 
configurations, but it still remains to demonstrate the translation of diagrams 
into the matching analytic expressions. Consider, say, the six-point $\mc N = 3$ 
and a seven-point $\mc N = 2$ MHV tree-level amplitudes of 
Fig.~\ref{SIXSEVENPOINT}. These diagrams were constructed simply by selecting 
two states from the $\Psi$ superfield and then patching up using $\Phi$
respecting that the total number of indices should be $2\mc N$. Referring to 
Fig.~\ref{INDEXRULES} we almost effortlessly find
\begin{align}
A_6^{\mathrm{tree}}(\hat 1^-_{g^{abc}}\hat 2^{\,}_{s^a}
3^{\,}_{s^{bc}}4^+_{g}5^+_{g}6^+_{g}) &=
i\frac{\avg{12}\avg{q_{1a}q_{2a}}\avg{q_{1b}q_{3b}}\avg{q_{1c}q_{3c}}}{
\prod_{i=1}^6\avg{i(i+1)}} \;, \\
A_7^{\mathrm{tree}}(\hat 1^-_{f^a}2^+_{f^a}\hat 3^-_{f^b}
4^+_{f^b}5^+_{g}6^+_{g}7^+_{g}) &=
i\frac{\avg{13}^2\avg{q_{1a}q_{2a}}\avg{q_{3b}q_{4b}}}{
\prod_{i=1}^7\avg{i(i+1)}} \;.
\end{align}

% N = 2/3

\section{Supersymmetric Sums in Unitarity Cuts}
Despite being formulated for tree-level amplitudes, the idea of generating
functions and their diagrammatic representation fits excellently with
evaluation of loop amplitudes using the generalized unitarity cut method
\cite{Bern:2009xq}. Indeed, the required supersymmetric sum over all
possible on-shell states propagating in the intermediate channels may be
realized as Grassmann integration of superamplitudes, while the flow of
$R$-symmetry charges is captured by index diagrams.

As an example we study a one-loop unitarity cut of a four-point amplitude with
external gluons, and carry out the supersum for the $\mc N = 2$ case.
For a more thorough description of unitarity cuts see 
\cite{Bern:1994cg,Bern:1994zx} and later developments \cite{Bern:1997sc,
Bern:2004ky,Bern:2004cz,Britto:2004nc,Bern:2007ct}. Evaluation of supersums has been discussed in 
\cite{Bern:2009xq,Elvang:2011fx,ArkaniHamed:2008gz,Bianchi:2008pu,
Elvang:2008na,Drummond:2008bq}. Here we follow the strategy of 
\cite{Bern:2009xq}.

\begin{figure}[!h]
\centering
\includegraphics[scale=0.6]{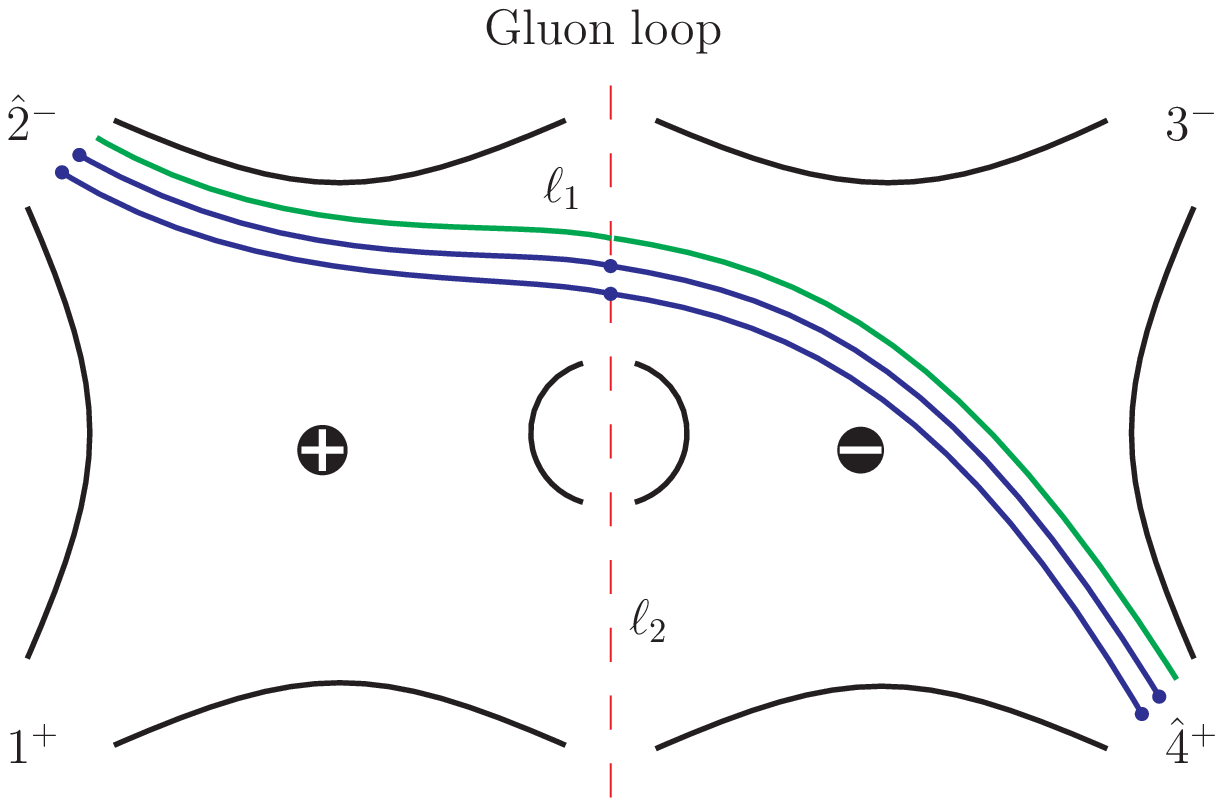}
\includegraphics[scale=0.6]{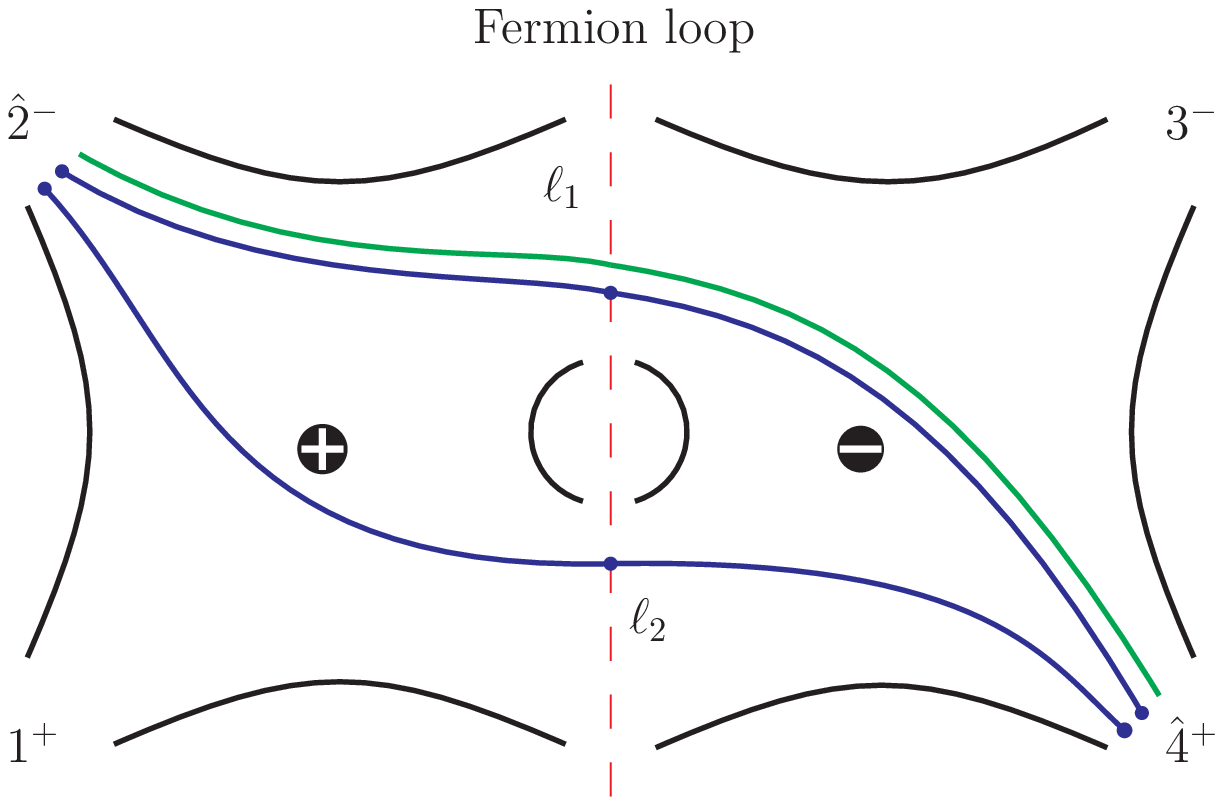}
\caption{\label{LOOPDIAGRAMS}
The left and right index diagrams should respectively illustrate internal gluon
and fermion contributions in a unitarity cut of the four-point one-loop
amplitude. The cut marked by the dashed red line splits the amplitude into
$\MHV$ and $\gMHV$ parts. Again the green sector line accounts for reduced 
supersymmetry. Horizontal flips of these diagrams and two additional diagrams 
representing internal scalars are very easy to draw, but are left out here.
}
\end{figure}

The supersum in question receives in total eight contributions, 
two with internal gluons, four with a fermion loop and two having scalars. 
Fig.~\ref{LOOPDIAGRAMS} provides two of these diagrams, and using the rules given 
in Fig.~\ref{INDEXRULES}, we find that their numerator part translate into
\begin{align}
\avg{q_{2a}q_{\ell_1 a}}\Avg{\tilde{q}_{\ell_1}^a\tilde{q}_4^a}
\avg{q_{2b}q_{\ell_1 b}}\Avg{\tilde{q}_{\ell_1}^b\tilde{q}_4^b}
\aAvg{2}{\ell_1}{4}^2 \quad
\mbox{and} \quad
\avg{q_{2a}q_{\ell_1 a}}\Avg{\tilde{q}_{\ell_1}^a\tilde{q}_4^a}
\avg{q_{2b}q_{\ell_2 b}}\Avg{\tilde{q}_{\ell_2}^b\tilde{q}_4^b}
\aAvg{2}{\ell_1}{4}^2,
\end{align}
where $\avg{ij}\Avg{jk} = \aAvg{i}{j}{k}$ for shorthand.

By inspection all eight diagrams have the same relative sign because the
corresponding Grassmann expressions are ordered equivalently. Therefore the
$\eta$'s may just be suppressed, and the numerator of the supersum thus becomes
\begin{align}
\aAvg{2}{\ell_1}{4}^4+\aAvg{2}{\ell_2}{4}^4+
2\aAvg{2}{\ell_1}{4}^3\aAvg{2}{\ell_2}{4}+
2\aAvg{2}{\ell_1}{4}\aAvg{2}{\ell_2}{4}^3+
2\aAvg{2}{\ell_1}{4}^2\aAvg{2}{\ell_2}{4}^2 \nonumber \\ =
(\aAvg{2}{\ell_1}{4}+\aAvg{2}{\ell_2}{4})^2
\times(\aAvg{2}{\ell_1}{4}^2+\aAvg{2}{\ell_2}{4}^2) \;.
\end{align}

An interesting pattern emerges when comparing the supersums for $\mc
N = 1,2,3,4$. Adding the diagrams in the right combinations dictated by the
different supermultiplets we see that the displayed resummation property 
is a common feature. Indeed, for maximal supersymmetry the supersum takes the 
very compact form $(\aAvg{2}{\ell_1}{4}+\aAvg{2}{\ell_2}{4})^4$, while for 
$\mc N < 4$
\begin{align}
(\aAvg{2}{\ell_1}{4}+\aAvg{2}{\ell_2}{4})^{\mc N}
\times(\aAvg{2}{\ell_1}{4}^{4-\mc N}+\aAvg{2}{\ell_2}{4}^{4-\mc N}) \;,
\end{align}
in agreement with \cite{Bern:2009xq}.

Cuts and supersums are by no means limited to MHV amplitudes. Non-MHV
amplitudes may be generated from MHV ones using the MHV vertex construction
\cite{Cachazo:2004kj,Risager:2005vk} as addressed in
\cite{Bern:2009xq,Bianchi:2008pu,Elvang:2008na,Georgiou:2004by},
and the MHV techniques thus apply.

\section{Conclusion}
In this paper we have investigated the superspace formalism of $\mc N = 4$ 
super Yang-Mills theory and its extension to situations with less than maximal
supersymmetry. We have written a general $\mc N$-dependent form of the MHV 
generating function at tree-level. More importantly, an extension to $\mc N$-fold
supersymmetry of a recent scheme for representing $\mc N = 4$ superamplitudes 
diagrammatically was presented. Although simple, it is nice to see
that the technique carries over from $\mc N = 4$. With this diagrammatic 
prescription it is extremely easy to memorize any super Yang-Mills scattering 
amplitude and translate it into analytic expressions, as we illustrated through 
several examples at both tree-level and one-loop. 

We will leave the exploration of multiloop unitarity cuts of non-MHV amplitudes 
utilizing the full content of the various $\mc N < 4$ multiplets for external 
states for future work. Another direction could be to study diagrams and
supersums in super Yang-Mills theory coupled to matter.

\begin{acknowledgments}
The author would like to thank Poul H. Damgaard and N. E. J. Bjerrum-Bohr for
helpful discussions.
\end{acknowledgments}

\end{document}